\documentclass[conference]{IEEEtran}
\IEEEoverridecommandlockouts
\usepackage[letterpaper, left=0.75in, right=0.75in, bottom=0.95in, top=0.7in]{geometry}
\usepackage[utf8]{inputenc}
\usepackage{subfigure,times,graphics,epsfig,amsmath,xspace,endnotes,pifont,multirow,rotating,listings,amssymb,algorithmic,color,caption,nicefrac,adjustbox,todonotes,tabularx,mathtools, algorithmic,algorithm}
\usepackage{graphicx}
\usepackage{tabularx}
\newcolumntype{L}[1]{>{\raggedright\arraybackslash}p{#1}} 
\newcolumntype{C}[1]{>{\centering\arraybackslash}p{#1}} 
\newcolumntype{R}[1]{>{\raggedleft\arraybackslash}p{#1}} %

\usepackage{amsthm}

\newtheorem{thm}{Theorem}
\newtheorem{lemma}[thm]{Lemma}

\newtheorem{defn}[thm]{Definition}
\usepackage{amsmath}
\usepackage{amssymb}
\usepackage{accents}

\usepackage{dsfont}
\usepackage{amsmath}
\usepackage{amssymb}

\begin{document}
%
\title{Controller Placement in SDN-enabled 5G Satellite-Terrestrial Networks}


\author{\IEEEauthorblockN{Nariman Torkzaban, and John S. Baras}
\IEEEauthorblockA{\textit{Department of Electrical and Computer Engineering}\\
\textit{\& Institute for Systems Research}\\
\textit{University of Maryland, College Park, Maryland, USA}\\
Email: \{narimant, baras\}@umd.edu}
}


\maketitle

\begin{abstract}
SDN-enabled Integrated satellite-terrestrial networks (ISTNs), can provide several advantages including global seamless coverage, high reliability, low latency, etc. and can be a key enabler towards next generation networks. To deal with the complexity of the control and management of the integrated network, leveraging the concept of software-defined networking (SDN) will be helpful. In this regard, the SDN controller placement problem in SDN-enabled ISTNs becomes of paramount importance.

In this paper, we formulate an optimization problem for the SDN controller placement with the objective of minimizing the average failure probability of SDN control paths to ensure the SDN switches receive the instructions in the most reliable fashion. Simultaneously, we aim at deploying the SDN controllers close to the satellite gateways to ensure the connection between the two layers occurs with the lowest latency. We first model the problem as a mixed integer linear program (MILP). To reduce the time complexity of the MILP model, we use submodular optimization techniques to generate near-optimal solutions in a time-efficient manner. Finally, we verify the effectiveness of our approach by means of simulation, showing that the approximation method results in a reasonable optimality gap with respect to the exact MILP solution. 


\end{abstract}



%
\IEEEpeerreviewmaketitle

\section{Introduction}

Over the past few years, it has become evident that towards achieving the key promises of 5G, it is essential to take advantage of the full capacity of all communications types \& segments (e.g. terrestrial, aerial, and space) as well as supporting technologies (e.g. SDN) simultaneously, otherwise the traditional stand-alone terrestrial networks will fail to achieve the key projected promises.
Integrating satellite and terrestrial networks can provide several advantages.

The integrated satellite-terrestrial network (ISTN) offers potential benefits which are not possible otherwise, including global coverage, low latency and high reliability. In particular, satellites can replace, extend, or complement the terrestrial networks, in rural, and hard-to-reach areas, or where the existence of communications infrastructure is costly or even infeasible in use-cases such as mountains and marine communications. Furthermore, satellites can offload the terrestrial networks by accommodating the delay-insensitive applications, allowing the terrestrial segment to survive when there is a surge in the traffic load. Finally, due to their global coverage, satellites can provide a reliable and seamless back-haul for aerial segment, and also the monitoring and control applications for IoT, vehicular networks, etc. On the other hand, despite providing the above advantages, new challenges are also introduced in the integrated network due to the limitations of the two different layers, including but not limited to complicated end-to-end resource provisioning due to the additional resource constraints, high control complexity due to the different dynamics of each segment, non-unified interfaces between the layers, etc., which significantly impact the decisions regarding traffic routing, spectrum allocation, mobility management, QoS and traffic management, etc. These challenges, together with the diversity of 5G use-cases with large-scale applications, highlights the importance of a unified management and control structure, and a dynamic resource allocation policy which are both scalable and flexible enough to handle the increasing complexity. The key to address these issues is the concept of software-defined networking (SDN). SDN allows for separating the control logic of the network from its forwarding logic and realizes a centralized management policy. This not only allows for a simple realization of the forwarding layer, but also paves the way for dynamic configuration of control and management policies.

Towards realizing the SDN-enabled ISTN, important optimization problems arise immediately; i) In the architectures concerned with GSO satellites, due to high delivered throughput per satellite, large number of gateways are required; sometimes exceeding a couple of dozens. ii) Moreover, once the gateway deployment policy is decided, it is of paramount importance to develop a smart and adaptive mechanism to handle the user hand-overs between the gateways or LEO satellites, traffic routing, load balancing, etc. Due to their abstract view of the network, SDN controllers are the best fit for this purpose. Thus, it becomes essential to formulate an optimization problem for deciding the minimum number of gateways and SDN controllers and their optimal location within the ISTN. Our approach to solve this problem is sequential. In \cite{ICC2020}, we formulated the satellite gateway placement problem while also optimizing the traffic routing. In this paper, we model and solve the SDN controller placement problem in 5G-satellite hybrid networks. In particular,

    i) Having provided the placement of satellite gateways, we formulate and solve the SDN controller placement problem in 5G-satellite hybrid networks, with the goal of maximizing the average control path reliability, and minimizing the controller-to-gateway latency.
    
    ii) We formulate the problem as a MILP and use CPLEX commercial solver to generate exact solutions for small-scale networks. For large networks, we use \textit{submodular} \textit{optimization}-based techniques to generate near-optimal solutions in a time-efficient manner.
    
    iii) We conduct extensive experimental tests to evaluate the performance of the provided methods and algorithms. We use publicly available real-world scenarios and various simulation settings for the performance evaluation tasks.

The remainder of the paper is organized as follows. Section~\ref{sec:desc} describes the network model and the problem description. In Section~\ref{sec:problem} we introduce the MILP formulation and its equivalent in the submodular optimization framework. Section \ref{sec:evaluation} presents our evaluation results, whereas Section \ref{sec:relatedwork} provides an overview of related work. Finally, in Section \ref{sec:conclusions}, we highlight our conclusions and discuss directions for future work.

\section{Network Model and Problem Description} 
\label{sec:desc}

\begin{figure}[t]
\begin{center}
\begin{minipage}[h]{0.489\textwidth}
\includegraphics[width=1\linewidth]{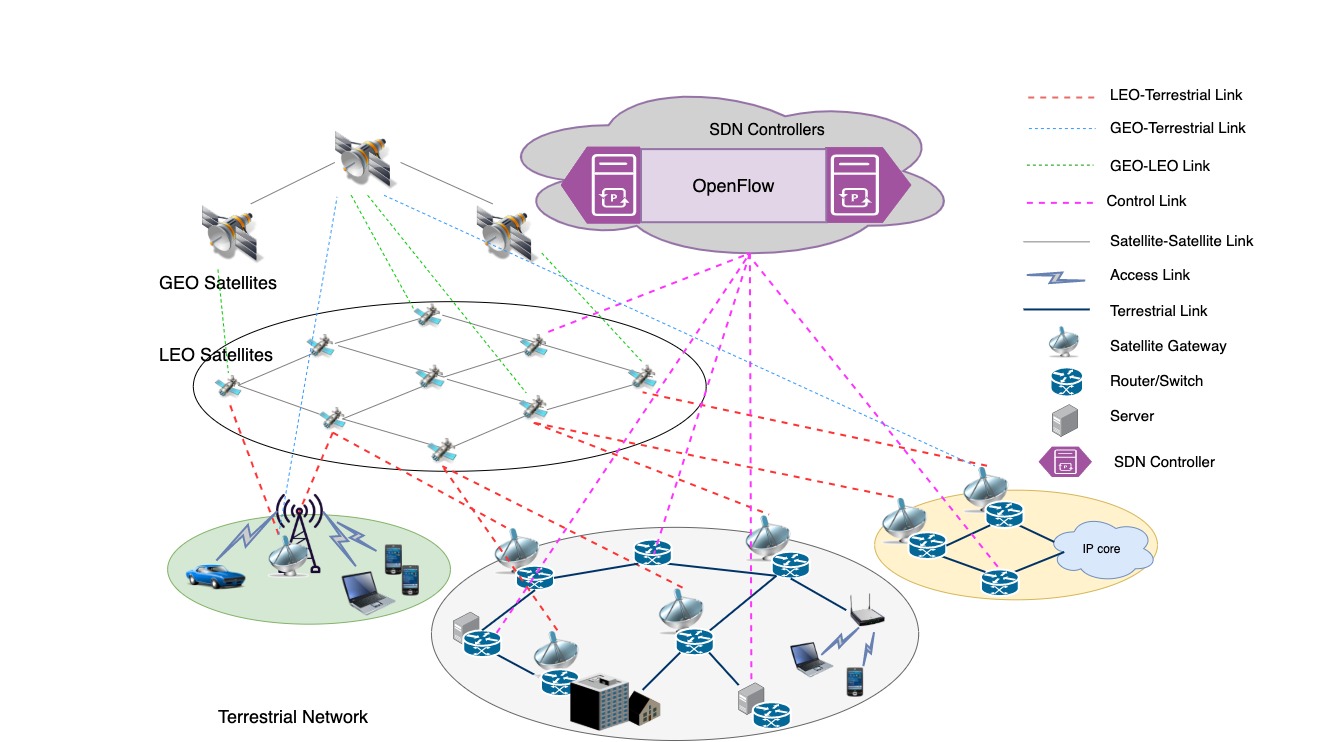}
\caption{ SDN-enabled 5G Terrestrial-Satellite Network}
\label{fig:scene}
\end{minipage}
\hspace{1em}
\end{center}
\end{figure}

We consider an SDN-enabled network architecture as in fig. \ref{fig:scene}, consisting of a \textit{control plane} and a \textit{data plane}. The control plane mainly consists of all the SDN controllers deployed on top of the physical hardware residing in the terrestrial or space segment within the hybrid network that will realize a logically centralized but physically distributed control scheme maintaining a global view of all portions of the network at all times. The data plane however, consists of the SDN-enabled switches mainly responsible for simply forwarding and receiving the traffic, leaving all the other management and control decisions  within all segments of the network (i.e. backhaul, core, access, etc.), to the control plane. 

Within the terrestrial segment of the network, the users deliver their traffic to the 5G core through the radio access network (RAN), i.e. small cells and gNBs, while the space segment mainly includes GEO satellites communicating with one another through laser link. The communication within the terrestrial segment relies on fiber links, whereas satellite gateways and relay nodes (RNs) facilitate the communication between ground and space layers. 

We consider an undirected graph $\mathcal{G} = (\mathcal{V}, \mathcal{E})$ to model the ground network where $(u,v) \in \mathcal{E} $ if nodes $u, v \in \mathcal{V}$ are directly connected. Let set $\mathcal{G}\in \mathcal{V}$ be the set of all those locations on the graph at which a gateway is placed. Moreover, let set $\mathcal{K} \subseteq \mathcal{V}$ be the set of all potential switches for hosting the SDN controllers. The sets $\mathcal{K}$ and $\mathcal{G}$ are not necessarily disjoint; therefore any terrestrial node $v \in \mathcal{V}$ may be an initial demand point, host a gateway or an SDN controller, or multiple of these.

The main functionality of the SDN controllers in the described architecture is to perform the routing decisions, manage the traffic handover among the satellite switches and gateways, and provide the necessary instructions to the SDN-enabled network switches. This will render the network switches as simple programmable forwarding devices. Given the central role of the control plane in the operation of the above-mentioned network, it is significantly important to ensure that the instruction paths between the SDN-enabled switches and the controllers are secure and reliable enough, because any failure or inefficiency in the nodes or links in the instruction path will block the network switches from receiving the accurate instructions and therefore may drastically impact the performance of the network. Moreover, given that the satellite gateways are the primary connection relay between the two network segments, it is important to ensure that the SDN controllers are deployed as close as possible to the gateways to ensure the instructions reach the space layer on time. To this end, we formulate the SDN controller placement in 5G satellite-terrestrial networks as an optimization problem with the objective of jointly minimizing the average controller-to-gateway latency and the average \textit{control path} error rate while being provided the gateway placement policy apriori. The optimal solution  outputs a set $\mathcal{K}^*$ of $k_1, k_2, \cdots, k_n$ locations for deploying the SDN controllers where $n$ is the number of selected controllers in $\mathcal{K}^*$, and the corresponding node-to-controller assignment policy.

We model the error rate $e_{ku}$ of a control path $P^{ku}$ between controller $k \in \mathcal{K}^*$ and terrestrial node $u \in \mathcal{V}$ as

\begin{equation}
    e_{ku} = 1 - (\prod _{e \in P_{ku}} {(1 - P_e)} \prod_{v \in P_{ku} }{ (1 - P_v)})
    \quad \forall k \in \mathcal{K}, u \in \mathcal{V} \label{error_c}
\end{equation} 

where $P_e$ and $P_v$ are the failure probability of edge $e$ and node $v$ belonging to the control path in respective order. 


\section{Problem Formulation}
\label{sec:problem}

We propose a solution to the SDN controller placement problem leveraging the concept of supermodularity of the cost function. We first formulate the problem as an integer program (IP), and then prove the supermodularity of the objective function. Then we invoke a linear-time algorithm with theoretical performance guarantee from the submodular optimization literature to generate near-optimal solution to the IP model. 

Let us define the following decision variables: 
\begin{itemize}
    \item The set of binary placement decision variables $\textbf{x}$ where $x_k = 1$, if a controller is placed at node $k \in \mathcal{K}$. 
    \item The set of binary assignment decision variables $\textbf{y}$ where $y_{kv} = 1$, if node $v \in \mathcal{V}$ is assigned to the controller placed at node $k \in \mathcal{K}$.
\end{itemize}

Our objective is inspired by the discussion in section \ref{sec:desc}, and therefore consists of two parts concerned with (i) Minimizing the average controller-to-gateway latency, and (ii) Minimizing the average control path error rate (average control path reliability maximization).

Let $W^c (x,y)$ be the first term of the objective function determining the average controller-to-gateway latency.  i.e.

\begin{equation}
    W^c (x,y) = \sum_{k \in \mathcal{K}}{d_k}{x_k}
\end{equation}

where $d_k$ is the amount of latency corresponding to the shortest path between location $k$ and the set of deployed gateways.

The second term of the objective function $W^r (x,y)$ determines the sum of failure probabilities of the control paths. 
\begin{equation}
    W^r (x,y) = \sum_{k \in \mathcal{K}}\sum_{v \in \mathcal{V}}{e_{kv}}{y_{kv}}
\end{equation}



Therefore, the objective function of the controller placement optimization problem will be 


\begin{equation}
    W(x,y) = \alpha W^c (x,y) +  W^r (x,y) \label{obj}
\end{equation}

where $\alpha >0$ adjusts the emphasis of the optimization problem on its two terms. 

The constraints of the optimization model are as follows:

No controllers can be placed at locations that are not a candidate for controller placement. i.e. 
\begin{equation}
    x_k = 0 \quad \forall k \in  \mathcal{V} \setminus \mathcal{K} \label{notplace}
\end{equation}
Each node has to be assigned to exactly one controller. Therefore, the following constraint is in place: 

\begin{equation}
    \sum_{k \in \mathcal{K}} y_{k v}=1 \quad \forall v \in \mathcal{V} \label{assC}
\end{equation}

Moreover the node-to-controller assignments must be valid, i.e. a node $v$ can only be assigned to the candidate node $k$ if a controller is placed at node $k$. Therefore, we have: 
\begin{equation}
    y_{k v} \leqslant x_{k} \quad \forall v \in \mathcal{V}, k \in \mathcal{K} \label{validc}
\end{equation}

The MILP formulation for the SDN controller placement will be as follows: 

 \begin{align}
 \text{Minimize} \quad &W(x, y) \label{kol}\\
 \text{subject to:}
& \quad \eqref{notplace}, \eqref{assC}, \eqref{validc}\\
& x_k \in \{0,1\}, \quad \forall k \in \mathcal{K}\\
& y_{kv} \in \{0,1\}, \quad \forall k \in \mathcal{K}, \quad \forall v \in \mathcal{V}
\end{align}

Next, we will provide a brief introduction to the concept of submodularity and submodular optimization. We will justify that our objective function belongs to the family of supermodular functions and on this ground we will invoke a heuristic method from the submodular optimization literature to solve our problem with provable theoretical guarantee.

\subsection{Sub (Super)-modular Optimization Methods}
\label{sec:submdis}

Let us start with the definition of submodular functions. 

\begin{defn}\textbf{Submodular Functions.} Let finite set  $G$ of elements be the ground set. Then A function $f: 2^{G} \rightarrow \mathbb{R}$ over the ground set is said to be submodular if for all subsets $A, B \subseteq G$, it holds that $$f(A)+f(B) \geq f(A \cup B)+f(A \cap B)$$

Equivalently, $f$ is said to be submodular if for all subsets $A, B \subseteq G$, 
with $A \subseteq B$ and every element $i \in G \backslash B$ it holds that:
$$
f(A \cup\{i\})-f(A) \geq f(B \cup\{i\})-f(B)
$$

\end{defn}

This intuitively means that for a submodular set function, adding an element to a subset will result in diminishing return with increasing the subset size. 
We also note that if for all subsets $A, B \subseteq G$, 
with $A \subseteq B$ it holds that $f(A) \leq f(B)$, then $f$ is called a \textit{monotone submodular function}. 

Moreover, it is worth noting that, if $f, g$ are submodular functions then $[f+g],[k f, k>0],[-f]$, are submodular, submodular, and supermodular in order.

Given the diminishing return property of the submodular functions, many utility functions can suitably fit in this class. Therefore, motivated by the natural application of submodularity property in real-world scenarios such as the welfare maximization, social networks, information gathering, feature selection, etc., optimization problems involving submodular/supermodular functions have developed a lot of interest among the research community. Especially, over the past decade, a lot of interesting methods have been proposed for approximately solving the submodular/supermodular optimization problems subject to a variety of constraints while providing acceptable optimality bounds. Within the SDN research community, several papers have used submodular optimization to model and solve multiple resource allocation problems \cite{cont}\cite{cont2}\cite{tass}.
Moreover, in \cite{tass2}, a very interesting taxonomy of such problems in the mobile edge computing (MEC) framework, along with an insightful discussion on their submodularity property is provided. 

In this paper, we illustrate that the model \eqref{kol} can be cast as a \textit{submodular optimization problem}. Then we will invoke an efficient linear-time method with theoretical approximation guarantee to solve the optimization problem. 

Following the approach in \cite{tass}, We first state a straightforward lemma showing that once the placement of the SDN controllers is fixed, then the optimal node-to-controller assignment policy can be obtained deterministically.

\begin{lemma}
Given a placement policy $\mathcal{X}$ for the SDN controllers, the assignment policy $\mathcal{Y}^r$ minimizing $W^r(x, y)$  can be uniquely determined as:

\begin{equation}
    \mathcal{Y}^r =  \{ (k_v, v): k_v = \arg\min{e_{kv}} \}\nonumber
\end{equation}
\label{lem1}
\end{lemma}

\textbf{Proof.}  When considering $W^r(x, y)$, each switch can simply choose the controller corresponding to the most reliable path regardless of the other switches. This clearly minimizes the average error rate for each node and therefore minimizes the total error rate. {$\blacksquare$}






This in effect means that the overall cost function can be modeled as a deterministic function of the SDN controller placement policy. Therefore, it makes sense to focus only on the optimization of the SDN controller placement policy. We will use this result in proving the following theorem which is central to our development. 

\begin{thm}
The cost function $W(x, y)$ is supermodular. \label{gate_min}
\end{thm}

\textbf{Proof.} Let $\mathcal{A}\subseteq\mathcal{B}\subseteq\mathcal{K}$ be two controller placement policies. Let $k \in \mathcal{K}\setminus\mathcal{B}$ be a new candidate node for deploying an SDN controller. We will show that each of the two terms of $W(x, y)$ is supermodular, therefore their summation is also supermodular. Let $M_i^{\mathcal{A}}(k)$ and $M_i^{\mathcal{B}}(k)$ be the amount of marginal addition to the objective function $W_i(x, y)$ when adding the new controller $k$, for $i \in \{c, r\}$. We have to show $\Delta M_i(k) = M_i^{\mathcal{B}}(k)$ - $M_i^{\mathcal{A}}(k)$ $\geq 0$

(i) Clearly, the marginal return to $W^c(x, y)$  by adding a new controller does not depend on the old placement policy and is equal to the unit cost of adding the new controller $d_k$. i.e. $\Delta M_c(k) = 0 $. Therefore, $W^c(x, y)$ is a modular function, and hence is supermodular by the definition.  

(ii) To show the supermodularity of $W^r(x, y)$, note that it follows from lemma \ref{lem1}, that $$ \forall v \in \mathcal{V}: e_{{k^\mathcal{A}_v}v} \geq e_{{k^\mathcal{B}_v}v}$$

Moreover, denote by $\sigma(\mathcal{V})$ the set of all those nodes that can increase the reliability of their control path by switching from ${k^\mathcal{B}_v}$ to the newly introduced controller $k$, i.e.
\begin{equation}
    \sigma(\mathcal{V}) = \{v \in \mathcal{V}: e_{kv} \leq e_{{k^\mathcal{B}_v}v}\} \label{changset}
\end{equation}

We will then have:

\begin{align}
    &\Delta M_c = \sum_{v \in \mathcal{V}}{\min(0, (e_{kv} - e_{k^{\mathcal{B}}_vv}))}\nonumber \\& - \sum_{v \in \mathcal{V}}{\min(0, (e_{kv} - e_{k^{\mathcal{A}}_vv}))}\label{mmmmm}\\
     &\geq \sum_{v \in \sigma(\mathcal{V})}{ (e_{kv} - e_{k^{\mathcal{B}}_vv})} - \sum_{v \in \sigma(\mathcal{V})}{ (e_{kv} - e_{k^{\mathcal{A}}_vv})} \label{vasat} \\
     &=\sum_{v \in \sigma(\mathcal{V})}{(e_{k^{\mathcal{B}}_vv} - e_{k^{\mathcal{A}}_vv})} \geq 0.
\end{align}

Therefore, $W^r(x, y)$ is supermodular. 

As a positively-weighted sum of two supermodular functions, $W(x,y)$ remains supermodular and the assertion in the theorem follows. {$\blacksquare$}

Now that we have shown the objective function belongs to the well-established class of supermodular functions, we are able to utilize the effective approaches for solving this type of problems. Particularly, let $\Bar{W}$ be an upper bound on $W(x,y)$. Therefore, it holds that $\Tilde{W}(x, y) = \Bar{W} - W(x,y)$ is a positive submodular function. Hence, equivalent to minimizing the supermodular function $W(x,y)$, we can maximize the submodular function $\Tilde{W}(x,y)$. This is in fact a typical approach for solving supermodular optimization problems. Several effective algorithms exist in the submodular optimization literature for approximately maximizing a positive submodular function with theoretical optimality gap guarantee in a time-efficient manner. In this paper, we will use a simple randomized linear-time $(1/2)$-approximation method to solve our problem. The $(1/2)$ is proved to be tight. 

\begin{thm}[\cite{focs}]
 There exists a $1/2$-approximation randomized greedy algorithm for maximizing a non-negative submodular function, which runs in linear time. 
 \label{yedovvom}
\end{thm} 

Algorithm \ref{alg1} shows the pseudo-code of how  the randomized method noted in theorem \ref{yedovvom}, will work in our settings. 

\begin{algorithm}
 \caption{(1/2)-approximation greedy algorithm}
 \label{alg1}
 \begin{algorithmic}[1]
  \renewcommand{\algorithmicrequire}{\textbf{Input:}}
  \renewcommand{\algorithmicensure}{\textbf{Output:}}
  \REQUIRE $\Tilde{W} : 2^ \mathcal{K} \rightarrow \mathds{R}_+$
  \ENSURE $(\Bar{X}, \Tilde{W} (\Bar{X}))$
 \STATE Initialize \quad $\underaccent{\bar}X = \emptyset , \quad \Bar{X} = \{1\}^{| \mathcal{K}|} $
 \STATE \textbf{for $k \in  \mathcal{K}:  $}
 \STATE \quad $\Bar{\Delta} = max(\Tilde{W} (\Bar{X}) - \Tilde{W} (\Bar{X}\setminus \{i\}), 0)$
 \STATE \quad $\underaccent{\bar}\Delta = max( \Tilde{W} (\underaccent{\bar}{X}) - \Tilde{W} (\underaccent{\bar}{X}\cup \{i\}), 0)$
 \STATE \quad \textbf{Set} $\quad  \underaccent{\bar}{X} = \underaccent{\bar}{X}\cup \{i\}$ with probability $\frac{\underaccent{\bar}\Delta}{(\underaccent{\bar}\Delta + \Bar{\Delta})}$
 \STATE \quad otherwise
 \STATE \quad \textbf{Set} $\quad \Bar{X} = \Bar{X}\setminus \{i\}$ 
 \STATE \textbf{end}
 \RETURN $(\Bar{X}, \Tilde{W} (\Bar{X}))$
 \end{algorithmic} 
 \end{algorithm}

 The algorithm starts by taking two extreme cases and then  decides on the placement of a controller at each location in an iterative fashion. Before $i$th iteration begins, a controller is present in $i$th location by the policy $\Bar{X}$ and absent by the policy $\underaccent{\bar}X$. The algorithm computes the contribution of the inclusion/exclusion of a controller in $i$th location, and makes a randomized choice accordingly. After iteration $i$ ends both the policies agree on the inclusion/exclusion of a controller at location $i$. Hence, when the execution of the algorithm finishes the two policies will be the same. The assignment policy can be computed according to lemma \ref{lem1}.

\section{Performance Evaluation}
\label{sec:evaluation}

In this section we will verify the effectiveness of our solution to the SDN controller placement by means of simulation. We will first describe the experiment setup and then the results.

\subsection{Setup and Parameters}
We consider real-world network topologies publicly available at the Internet Topology Zoo \cite{zoo} that are widely used in the literature of the related works. The complete list of topologies we have considered is listed in table \ref{topo}. In all the topologies the lengths of the links that are used in the shortest path algorithm are extracted and computed from \cite{zoo}. To generate the shortest paths between each pair of nodes we adopt an implementation of the Yen's algorithm as in \cite{yen}. Moreover, we adopt a similar approach to that of \cite{relcase} for computing the failure probabilities of the network components; We randomly generate the failure probabilities for terrestrial nodes, terrestrial links, and the satellite link, in $4$ different settings. Table \ref{failure} lists the range of failure probabilities in different cases for each network component. The placement of gateways is already given according to \cite{ICC2020}. To solve the MILP models we use CPLEX commercial solver, and conduct all the experiments on an Intel Xeon processor at 3.5 GHz and 16 GB of main memory. Each experiment is repeated $100$ times and the results are averaged. We will then compare the MILP model and the greedy algorithm. For more detailed performance analysis, the reader is referred to the long version of this work available at \cite{2103.08735}.

\begin{table}[t]
\centering
\caption{Network Topology Settings}
\begin{tabular}{||c c c||} 
 \hline
 Topology & Nodes & Links  \\ 
 \hline\hline
 Nsfnet & 13 & 15  \\
 Ans & 18 & 25  \\
 Agis & 25 & 32  \\
 Digex & 31 & 35  \\
 Chinanet & 42 & 86\\
 Tinet & 53 & 89\\
 \hline
\end{tabular}
\label{topo}

\vspace{4mm}
\caption{Failure Probability Settings}
\resizebox{\columnwidth}{!}{
\begin{tabular}{|c|c|c|c|}
\hline & $P_{v}$ terrestrial nodes & $P_{e}$ terrestrial links & $P_{e_{s} g}$ satellite links \\
\hline Case 1 & {[0,0.05]} & {[0,0.02]} & {[0,0.02]} \\
\hline Case 2 & {[0,0.06]} & {[0,0.04]} & {[0,0.03]} \\
\hline Case 3 & {[0,0.07]} & {[0,0.06]} & {[0,0.04]} \\
\hline Case 4 & {[0,0.08]} & {[0,0.08]} & {[0,0.05]} \\
\hline
\end{tabular}
}
\label{failure}
\end{table}

\subsection{Numerical Results}

\begin{figure*}[t]
\begin{center}
\begin{minipage}[h]{0.225\textwidth}
\includegraphics[width=1\linewidth]{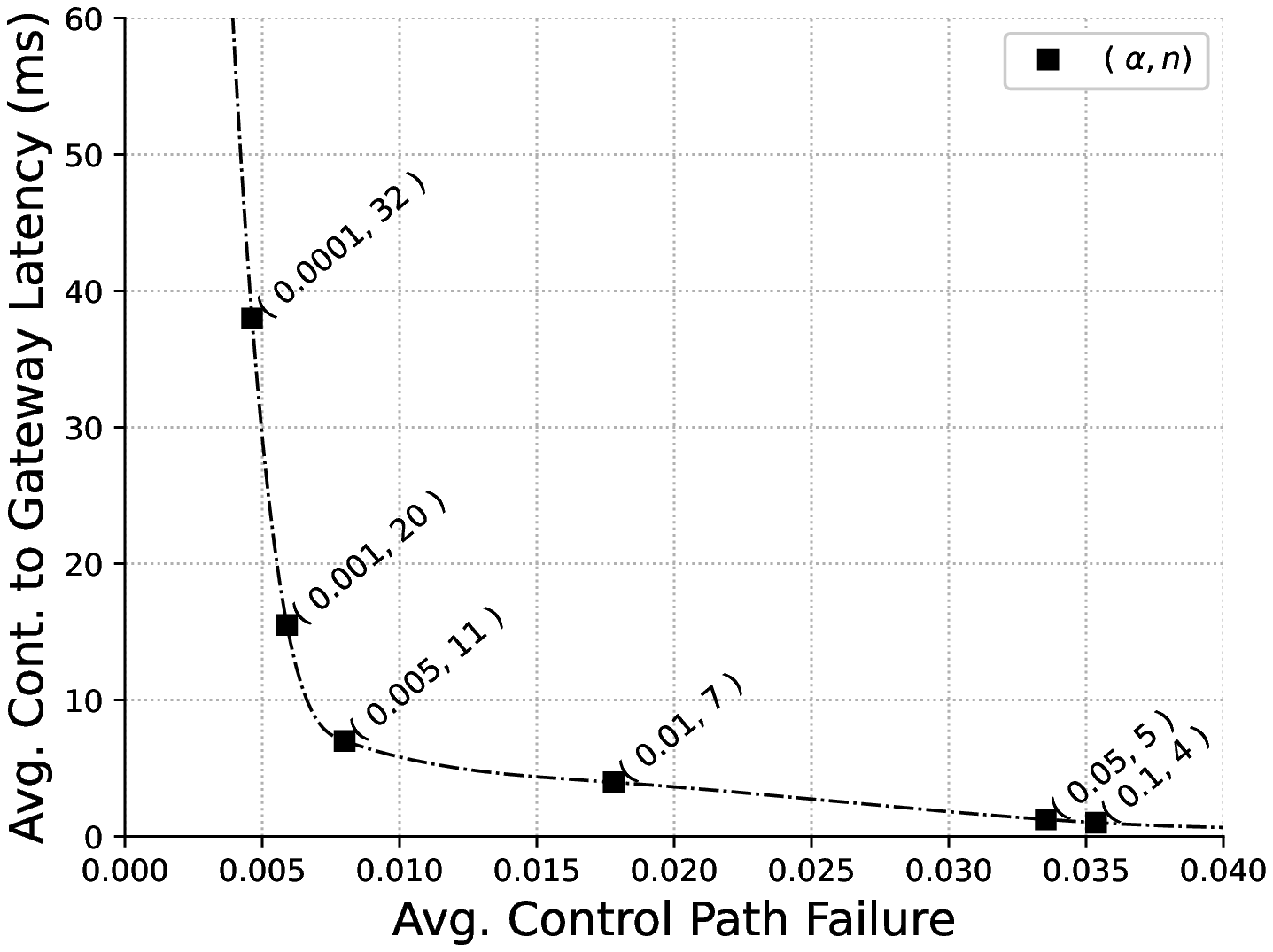}
\caption{ Effect of $\alpha$ on the objective balance}
\label{fig:tradeoff}
\end{minipage}
\hspace{1em}
\begin{minipage}[h]{0.225\textwidth}
\includegraphics[width=1\linewidth]{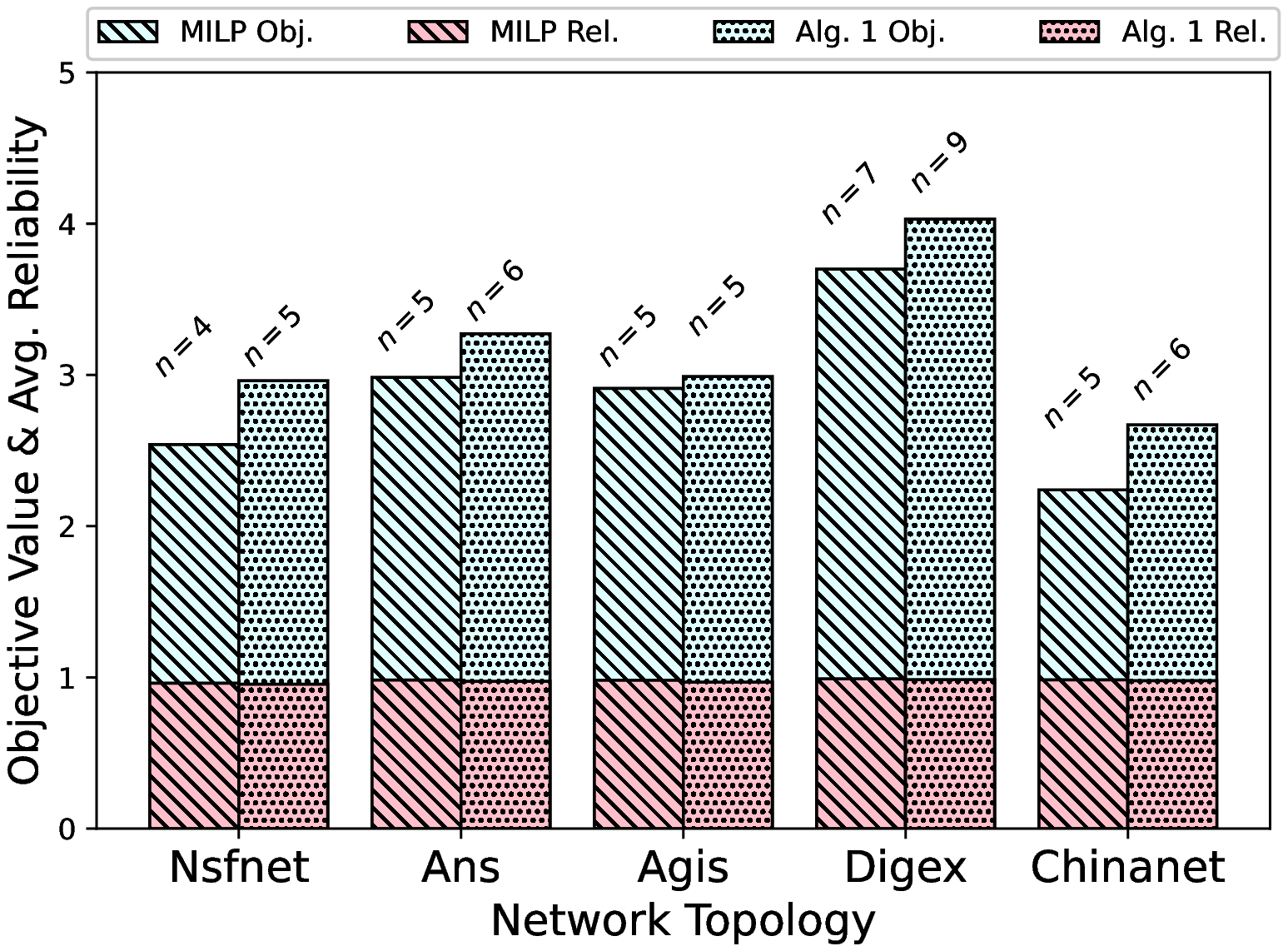}
\caption{ Comparison between MILP and Alg. $1$}
\label{fig:comp}
\end{minipage}
\hspace{1em}
\begin{minipage}[h]{0.225\textwidth}
\includegraphics[width=1\linewidth]{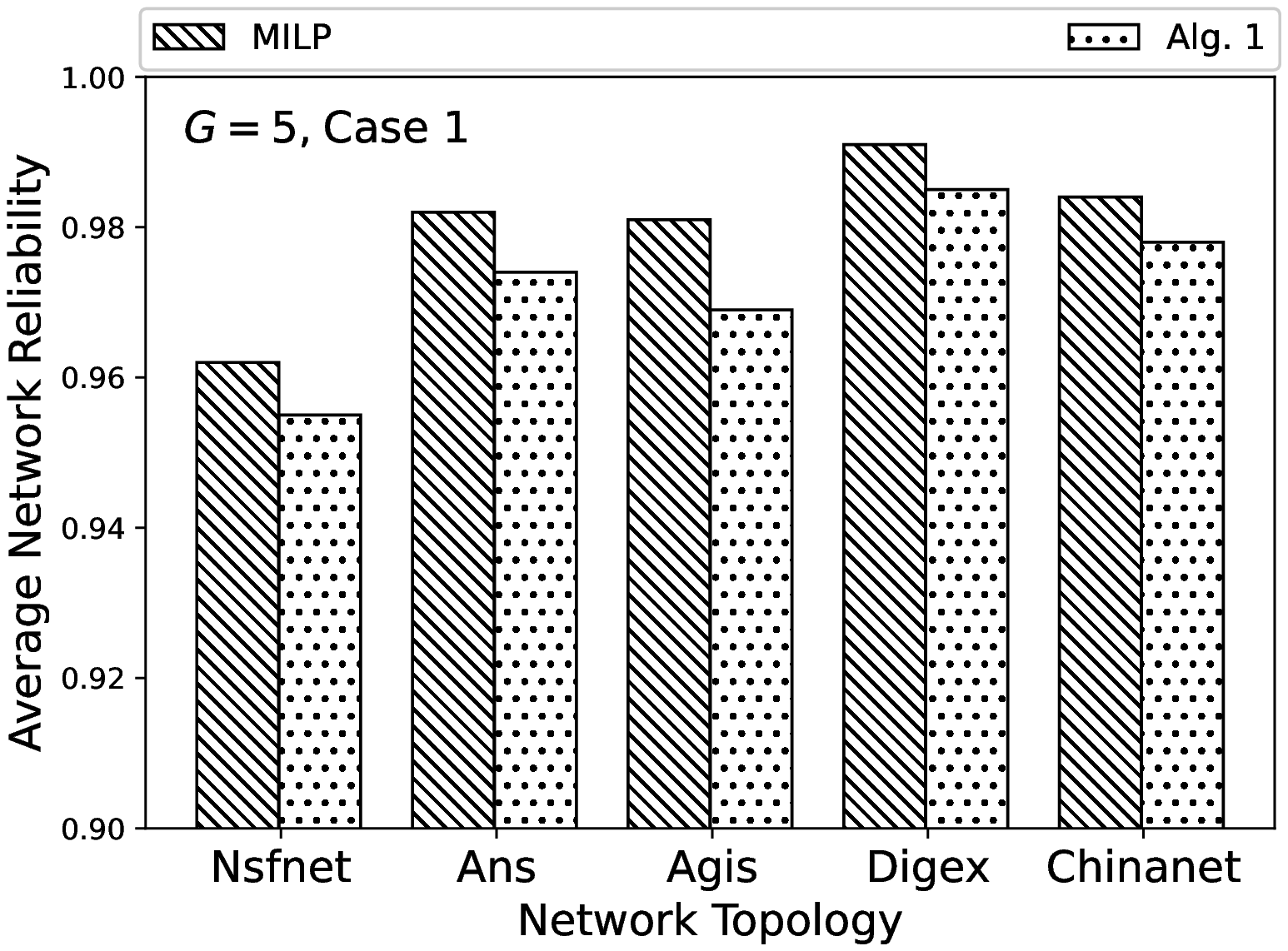}
\caption{ Avg. control path reliability: MILP \& Alg. $1$}
\label{fig:rel_comp}
\end{minipage}
\hspace{1em}
\begin{minipage}[h]{0.225\textwidth}
\includegraphics[width=1\linewidth]{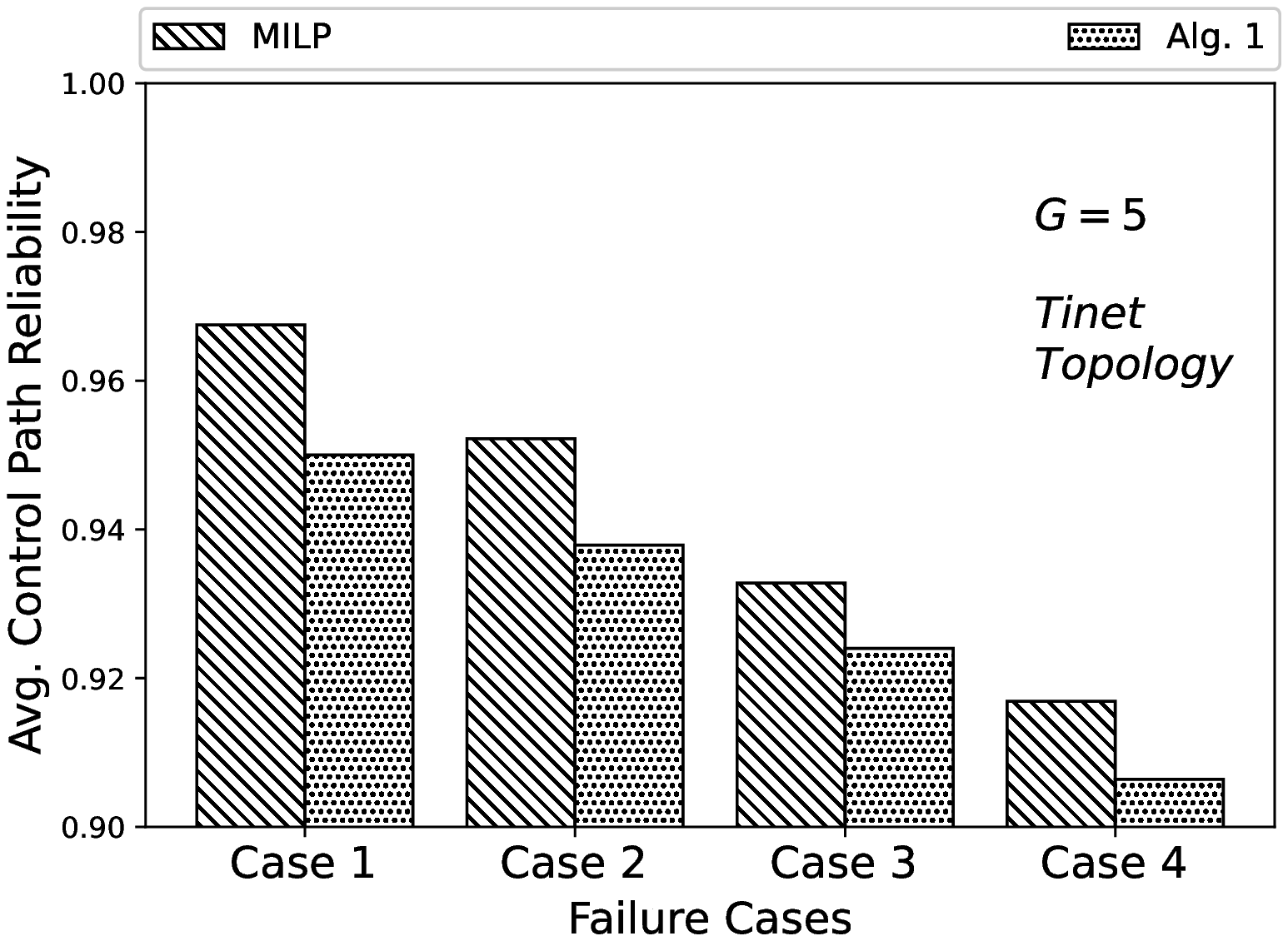}
\caption{ Avg. control path reliability in different cases}
\label{fig:falure}
\end{minipage}
\hspace{1em}

\end{center}
\end{figure*}

Figure \ref{fig:tradeoff} shows the impact of changing $\alpha$ on the tradeoff between the two terms of the objective function. In all cases $G = 5$ gateways are assumed to be deployed in the Tinet topology. For small values of $\alpha$, low emphasis is made on the first term of the objective but reliable control paths will be formed. As $\alpha$ increases the controllers move closer to the deployed gateways on average, at the expense of lower control path reliabilities. At each enumerated point, the corresponding value of $\alpha$ and the number of deployed SDN controllers are annotated. 

Figure \ref{fig:comp} presents the comparison between the performances of the exact MILP and the greedy randomized algorithms. The number of gateways is fixed at $G =  5$, and case $1$ is picked for failure probabilities. The resulting number of deployed controllers is annotated on each bar. The suboptimality gap remains between $12 \%$ in the objective function in all the cases while the greedy method runs much faster for larger networks. In fact, as we have shown in the long version of this paper available at \cite{2103.08735}, the submodular optimization method can run $100$ times in an amount of time that is negligible to running large-scale MILPs. 

Figure \ref{fig:rel_comp} shows in great detail, the performance of he two methods in terms of average control path reliability in accordance to figure \ref{fig:comp}. It is observed that the greedy method can maintain a close-to-optimal control path reliability maintaining at most $2\%$ gap in the average control path reliability compared to the exact MILP approach. 

Finally, figure \ref{fig:falure} shows how less reliable network components may impact the reliability of control paths. For this experiment $5$ gateways are assumed to be deployed in the Tinet topology. It is noteworthy that in all cases the performance of the greedy method remains within $2\%$ of that of the optimal one. 

The above results confirm that our model is effective and the approximate approach works reasonably well when comparing to the optimal solution. 

\section{Related Work}
\label{sec:relatedwork}
 With respect to SDN controller placement, the research works differ mostly in the location of placing the controllers each of which providing some benefits and some shortcomings. Some papers decide to place the controllers on the ground segment, some place them on the LEO SDN-enabled satellite switches \cite{wolf}\cite{wolf2}, and some on the GEO layer \cite{geoc}, while some other works propose hierarchical controller architectures comprising ground stations, LEO, MEO, and GEO-layer controllers \cite{hier}, \cite{hierm}, \cite{hier3}.  
 In \cite{hierm}, \cite{wolf}, and, \cite{wolf2}, the authors consider the controller placement problem in both static and dynamic modes, where in the former the controller placement and satellite-to-controller assignments remain unchanged, while in the latter the number of the controllers, their locations, and therefore the assignments vary with respect to change in demand and traffic pattern over time. Further in \cite{wolf}, and \cite{wolf2}, the flow setup time is adopted as a metric which makes the problem statement realistic as optimizing the flow setup time is a major concern in SDN-enabled networks.

The joint gateway deployment and controller placement in ISTN has also received an increasing attention over the past few years. The authors in \cite{scgp}, formulate a joint deployment of satellite gateways and SDN controllers to maximize the average reliability with hard constraints on user-to-satellite delay. They propose an iterative approach based on simulated annealing and clustering, where in each iteration first the current gateway placement policy and then the controller placement is updated towards the convergence. In \cite{part}, the exact same problem under similar settings and with similar objectives has been considered with the only difference that  the simulated annealing approach from \cite{scgp} is augmented with a portioning phase (separately w.r.t gateways \& controllers) to render several sub-problems of smaller size. Finally, in \cite{meta}, a number of meta-heuristic approaches namely, simulated annealing, double simulated annealing, and genetic algorithm for the same problem has been considered and their performance is compared. 
\section{Conclusions}
\label{sec:conclusions}
In this paper, we studied the SDN controller placement problem in ISTNs to jointly minimize the controller-to-gateway latency and maximize the control path reliability. We show that our objective function is supermodular and apply a heuristic approach from the submodular optimization community to generate near-optimal solutions for large-scale networks in a time-efficient manner. We verify the effectiveness of our approach by means of simulation and comparing our results to state of the art. 

Moreover, within the framework of 5G-Satellite integration, SDN controller placement for LEO constellations becomes more important due to the more frequent need for hand-offs between the satellite switches; and at the same time is more challenging due to the dynamically changing network topology and potentially large number of SDN controllers required in both the terrestrial and the space layer. This problem is among our future research directions.

\bibliographystyle{IEEEtran}
\bibliography{bibliography}
\end{document}